\documentclass[12pt, preprint]{aastex}

\newcommand{\merr}[3]{$#1_{#2}^{#3}$}
\newcommand{\cosec}{\mathop{\rm cosec}\nolimits}
\newcommand{\cind}[1]{\noindent{\bf #1\/:} }
\newcommand{\papern}[1]{Paper~II}

\slugcomment{AJ, in press}

\shorttitle{MS4 Sample. I}
\shortauthors{Burgess \& Hunstead}

\begin{document}

\title{The Molonglo Southern 4\,Jy Sample (MS4).  I.\\
    Definition}

\author{A.\ M.\ Burgess\altaffilmark{1}}
\email{annb@psych.usyd.edu.au}

\and

\author{R.\ W.\ Hunstead}
\affil{School of Physics, University of Sydney,
    NSW 2006, Australia}
\email{rwh@physics.usyd.edu.au}

\altaffiltext{1}{present address: School of Psychology, University of
Sydney, NSW 2006, Australia.}

\begin{abstract}
We have defined a complete sample of 228 southern radio sources at 408
MHz with integrated flux densities $S_{\rm 408} > 4.0$\,Jy, Galactic
latitude $|b|>10^{\circ}$ and declination $-85^{\circ} < \delta <
-30^{\circ}$.  The main finding survey used was the Molonglo Reference
Catalogue.  We describe in detail how the Molonglo Southern 4 Jy sample
(MS4) was assembled and its completeness assessed.  Sources in the
sample were imaged at 843\,MHz with the Molonglo Observatory Synthesis
Telescope to obtain positions accurate to about $1''$, as well as flux
densities and angular sizes; follow-up radio and optical observations
are presented in Paper II.  Radio spectra for the MS4 have been compiled
from the literature and used to estimate flux densities at 178\,MHz.
The strong-source subset of MS4, with $S_{\rm 178} > 10.9$\,Jy (SMS4),
provides a southern sample closely equivalent to the well-studied northern
3CRR sample.  Comparison of SMS4 with 3CRR shows a reassuring similarity
in source density and median flux density between the two samples.
\end{abstract}


~~~~
\clearpage

\section{INTRODUCTION}

Well defined, complete surveys are essential starting points for
learning about the global characteristics of astronomical sources.  At
radio frequencies, the easiest way to select a large, well defined
sample is to choose all objects brighter than a given flux density at a
given observing frequency.  Such a sample has the advantage of
containing the objects with the highest signal-to-noise ratio.  Because
any one finding survey is likely to be affected either by confusion or
by over-resolution for some of the sources, it is often necessary to
use more than one survey to define a flux density-limited sample.  Even
then the sample may not be complete, but may, for example, be missing
sources (or portions of sources) of low surface brightness or very
large angular size.

One of the best studied radio samples in the past few decades has been
the northern hemisphere Third Cambridge Catalogue (3C; \citealp{3c})
and its revised versions the 3CR \citep{3cr} and 3CRR \citep{lrl}.
Being selected at low frequency (159 and later 178\,MHz), it contains
a large proportion of extended, steep-spectrum sources.  Because of
the high flux-density limit ($S_{\rm 178}$) of around 11\,Jy, the
sample contains a large proportion of very powerful sources.  The most
complete version, the 3CRR, is the one discussed in this paper.

As so many studies of high-power radio sources have been based on
this one sample, it is essential to have a comparison sample to test
whether the 3CRR is truly representative.  This paper is concerned with
the definition of such a comparison sample, selected at low
frequency to have properties similar to 3CRR.  Increasing the sky
coverage is more important at high flux densities where the total
number of sources is limited.  This sample was selected at 408\,MHz to
contain the $\sim 200$ brightest extragalactic objects south of $\delta
= -30^{\circ}$.

The sample is defined in Section~\ref{sec.sampdef}.  Follow-up
observations of the whole sample with the Molonglo Observatory Synthesis
Telescope at 843\,MHz are described in Section~\ref{sec.mostobs}.  These
observations have provided more accurate positions and angular sizes, as
well as flux densities in a gap in the radio spectrum.  The definition of
a strong subsample at 178\,MHz, for comparison with 3CRR, is described
in Section~\ref{sec.est178}.  In \papern\/ \citep{paper2} we present
follow-up imaging at 5\,GHz of the more compact sources with the Australia
Telescope Compact Array, together with optical identifications using the
UK Schmidt sky survey and CCD images from the Anglo-Australian Telescope.

\section{DEFINITION OF THE MS4 SAMPLE}
\label{sec.sampdef}

Our motivation was to generate a sample of strong southern radio
sources for comparison with (and extension of) the well studied
northern 3CRR sample.  For reference, Table~\ref{tab1.surveys} contains
a summary of southern finding surveys of strong sources.  Our
selection frequency of 408\,MHz (as opposed to 178\, MHz used for
3CRR) was chosen because it allowed the sources to be selected from
the Molonglo Reference Catalogue (MRC; \citealp{rad.L1}), the deepest
and most complete low-frequency survey of the southern sky.  The
flux-density limit of 4.0\,Jy was chosen to give a useful but manageable
number of sources, $\sim$200--250.  The sample is therefore called the
Molonglo Southern 4\,Jy sample, abbreviated as MS4.

The selection criteria for the MS4 sample are as follows: 

\begin{itemize}

\item [(i)] Declination $-85^{\circ} < \delta < -30^{\circ}$.

\item [(ii)] Flux density $S_{\rm 408} > 4.0$\,Jy.

\item [(iii)] Galactic latitude $|b| > 10^{\circ}$.

\item [(iv)] Not in the Magellanic Cloud regions.

\item [(v)] Not a known Galactic source.

\end{itemize}

The reason for excluding the small area south of $-85^{\circ}$ is that
the MRC is incomplete in this region.  At least one source would
otherwise be in the sample: PKS~B0349$-$88, with $S_{\rm 408} =
6.8$\,Jy \citep{rad.P1}.  The northern limit of $\delta = -30^{\circ}$
was chosen to preserve a reasonable beamshape for follow-up images
made with the east-west arrays of the Australia Telescope Compact
Array (ATCA) and Molonglo Observatory Synthesis Telescope (MOST).  For
the purposes of sample definition the original B1950 coordinates were
retained, but all positions are given for the J2000 equinox.

The Galactic Plane and Magellanic Cloud regions are excluded for two
reasons: the strong radio emission from local sources makes background
extragalactic objects hard to recognise, and the dust obscuration and
high star density make optical identifications difficult.  The
Magellanic Cloud regions are those defined by \citet{clouds.survey},
i.e., $00^{\rm h}\,25^{\rm m} < \alpha < 01^{\rm h}\,35^{\rm m},
-74^{\circ}40' < \delta < -70^{\circ}45'$ for the Small Magellanic
Cloud, and $04^{\rm h}\,46^{\rm m} < \alpha < 05^{\rm h}\,58^{\rm m},
-72^{\circ}20' < \delta < -65^{\circ}10'$ for the Large Magellanic
Cloud.  Examination of the 408\,MHz MC4 survey \citep{rad.C2} reveals
that the only sources in these regions with $S_{\rm 408} > 4.0$\,Jy
are associated with the Clouds rather than being background objects.
From the mean source density of MS4 (\S~\ref{sec.thesample}) we
would expect only two sources to lie in these regions, so a count of
zero is not surprising.

\subsection{The Molonglo Cross Telescope and the Molonglo Reference Catalogue}

The Molonglo Reference Catalogue (MRC) was compiled from data taken
between 1968 and 1978 with the Molonglo Cross Telescope.  This
telescope was a Mills Cross interferometer operating at 408\,MHz with
a bandwidth of 2.5\,MHz, and resolution of $2.67' \times 2.86' \sec
(\delta + 35^{\circ}.4)$ \citep{mills.cross}.  The flux-density
calibration \citep{posflux.molonglo} was based on the absolute scale
of \citet{flux408.wyllie} and the pencil-beam sensitivity curve
measured by \citet{opt.h5}.  The survey for the MRC covered the
declination range $-85^{\circ} \leq \delta \leq +18^{\circ}.5$, with
Galactic latitude $|b| > 3^{\circ}$.  The MRC contains 12141 sources
above its lower flux-density limit of $S_{\rm 408}=0.7$\,Jy, and
contains 7347 sources above its notional completeness level of 1\,Jy.
As well as positions and flux densities, the catalogue contains
cross-references to other catalogues, and some structural information
for extended sources.  For sources stronger than 4\,Jy the positional
errors are around $3''$ in $\alpha$ and $4''$ in $\delta$, and the
standard error in flux density is around 4\%.

In defining our sample, we used the MRC in preference to the two
previous extensive low-frequency surveys of the southern sky, the
Mills, Slee, and Hill (MSH) survey at 85.5\,MHz
\citep{msh1,rad.M1,rad.M2} and the Parkes 408\,MHz catalogues
\citep{rad.B1,rad.P1}.  Compared with the MRC, the MSH and Parkes
catalogues have less uniform sky coverage, lower sensitivity, and
higher source confusion due to lower resolution.  Of the 228 sources
in the MS4 sample, 43 were not in the original Parkes survey: of
these, 35 were in the region of sky covered by that survey.  Because
of their much wider beams, the Parkes and MSH surveys were valuable in
providing integrated flux densities for strong but very extended
sources (such as the lobes of Centaurus A) which are substantially
resolved out by the Molonglo Cross.

\subsection{Initial Selection of the Sample}
\label{sec.initselec}

Initial selection of the MS4 sample from the MRC gave 215 sources
in the region of interest with $S_{\rm 408} > 4.0$\,Jy.  This list
was refined using the study of \citet{rad.J1}, hereafter referred to
as JM92.  They used the Molonglo Observatory Synthesis Telescope (MOST)
to image at 843\,MHz all MRC sources south of $-30^{\circ}$ which were
flagged as extended or multiple.  Their higher resolution MOST images
made it possible to distinguish between genuine extended sources, some
with more than one MRC entry, and close groups of unrelated sources.
Following \citet{paj.thesis}, two MRC sources were considered to be parts
of a single source if they were extended towards each other, or if they
were connected at or above the 2\% contour level.  This latter criterion
corresponds to a separation of about $105''$ for two unresolved sources.
For instance, we excluded MRC~B0230$-$666, with $S_{\rm 408} = 4.39$\,Jy,
from our sample because JM92 show it is composed of two separate compact
sources, each well below our 4\,Jy limit.  MRC~B1459$-$417, identified
with the Galactic supernova remnant SNR\,1006 \citep{snr.1006}, was also
excluded.  After excluding these two sources, and counting 5 sources
with multiple MRC entries only once, our sample contained 208 sources.
This was, however, still incomplete, because of the underestimation of
flux densities of extended sources in the MRC.

The MRC flux densities were calculated using a point-source fitting
algorithm which is reliable for sources with angular extent less than
$\sim1'$.  For larger angular sizes the fitted value underestimates
the integrated flux density, so it was necessary to estimate the
integrated $S_{\rm 408}$ for each extended source in the region of
interest.  We first obtained a candidate list of 178 sources with MRC
flux densities between 1 and 4\,Jy, and flagged as extended or
multiple; weaker sources were ignored as the MRC is incomplete below
1\,Jy.  Of the 178 sources, 97 were found by JM92 to have largest
angular size $> 1'$.  We searched the literature for flux-density
measurements at 408\,MHz for these 97 sources.  If such measurements
were unavailable, we used the radio spectrum or unpublished Molonglo
408\,MHz data.  As a result of these estimates, described in the
following section, 20 extended sources were added to the sample,
bringing the total number of sources to 228.

\subsection{Flux Densities of Extended Sources}
\label{sec.extflux}

\subsubsection{Published Flux Densities at 408\,MHz}

408\,MHz flux densities for extended southern sources are available
from either the 64\,m Parkes Telescope or the Molonglo Cross.  The
Molonglo papers used the absolute flux-density scale of
\citet{flux408.wyllie}, whereas the Parkes catalogues used the scale
of \citet{flux408.ckl}, which was nominally about 10\% lower.
However, \citet{rad.W6}
found a mean ratio $\langle$Molonglo/Parkes$\rangle$ $= 1.000 \pm
0.015$ and we have therefore not made any adjustments to the Parkes
flux densities.

The main Parkes 408\,MHz data were collected as part of the original
Parkes surveys \citep{rad.B1,rad.P1}.  The Parkes beam size of $48'$
often led to overestimation of flux densities because of confusing
sources in the beam.  We therefore used these flux densities only if
estimates via the radio spectrum (\S~\ref{sec.radspec}) were
unavailable, or for strong sources where confusion was unlikely to be
severe.  Parkes 408\,MHz flux densities were used to exclude four
sources from the sample.  Few of the remaining candidates were in the
Parkes catalogue, as they lay well below its stated completeness limit
of 4\,Jy.

The two main sets of Molonglo Cross data in the region of interest were
those of \citet{rad.S7}, hereafter referred to as SM75, and the MC4
survey of the Magellanic Cloud regions \citep{rad.C2}.  SM75 estimated
integrated 408 MHz flux densities for 116 sources found to be extended
with Parkes or Molonglo.  A total of 17 candidates from our list were
in their survey, of which 11 were listed as having $S_{\rm 408} >
4.0$\,Jy.  One of these candidates, MRC B2130$-$538, in the cluster
Abell\,3785, was subsequently excluded because a high-resolution image
\citep{haigh.thesis}, made with the Australia Telescope Compact Array
at 20\,cm, showed it to be two independent radio sources.  We also
included a source, PKS~B1400$-$33, which had $S_{\rm 408} > 4.0$\,Jy in
SM75, but did not appear in the MRC.

The MC4 finding survey covered the Magellanic Clouds, as well as
comparison regions in the declination range $-61^{\circ} \leq \delta
\leq -75^{\circ}$.  Of the 10 candidates observed in this survey, two
were found to have integrated flux density $S_{\rm 408} > 4.0$\,Jy.

Integrated flux densities at 408\,MHz measured with the Molonglo Cross
by J.~Rathmell \citep{rad.E1} were used to include MRC~B1056$-$360,
with $S_{\rm 408} = 4.09$\,Jy.

\subsubsection{Estimates from the Radio Spectrum}
\label{sec.radspec}

Because extended sources usually have power-law radio spectra, it is
often possible to estimate flux densities by linear interpolation or
extrapolation of a $\log{S_{\nu}}$: $\log{\nu}$ plot.  We used
flux-density measurements at other frequencies to estimate $S_{\rm
408}$ for sources without published integrated flux densities at
408\,MHz.  It was sometimes necessary to extrapolate to $S_{\rm
408}$, if data at lower frequencies were unavailable.  There will
be systematic error in these estimates both from missing flux density
for sources with large angular extent, and from confusion in
low-resolution observations, in particular those of MSH at 85.5\,MHz,
and Parkes at 408 and 1410\,MHz.

To quantify the error in the flux-density estimates, we compared flux
densities from SM75 with values interpolated or extrapolated from the
radio spectrum for 20 extended sources.  For 15 of these sources, the
estimate from the radio spectrum was larger than the SM75 value.  For
the 8 sources with largest angular size LAS $<7'$, the median
flux-density difference was 13\%, and for the 12 sources with LAS $\geq
7'$ the median flux-density difference was 24\%.  These systematic
differences are attributed to the limited surface brightness sensitivity
of the 3$'$ Molonglo Cross pencil beam.

On the basis of the radio spectra, two sources were added to the sample.

\subsubsection{Estimates from Unpublished Molonglo 408\,MHz Data}

The unpublished Molonglo Transit Catalogue, from an intermediate stage
in the compilation of the MRC, contains estimates of integrated flux
densities which we were able to compare with the corresponding values
in SM75.  The Transit Catalogue values were found to underestimate the
total flux density for sources of LAS $\gtrsim 6'$.  For sources with a
single MRC entry and LAS $<6'$ the estimates were reliable.  Of the 21
sources in this latter category, four were added to the sample.

After performing these various estimates of $S_{\rm 408}$, we were left
with 20 extended MRC sources to add to the MS4 sample.  They are listed in
Table~\ref{tab2.extra}.  Extended sources with integrated flux densities
just below the 4.0\,Jy cutoff are listed in Table~\ref{tab3.nearmiss}.

\subsection{Completeness}

\subsubsection{Systematic Errors}

In spite of the work described in Section~\ref{sec.extflux}, the MS4
sample may still be incomplete, due to very extended sources.  Since
the MRC is complete only for sources with fitted flux density
$>1$\,Jy, we may be missing some strong but diffuse sources with
surface brightness $<1$\, Jy/beam.  However, such sources should have
been catalogued in the lower resolution Parkes surveys, and a search
in the region of interest for non-MRC sources with $S_{\rm 408} >
4.0$\,Jy located only three objects.  Two of these, PKS~B1209$-$52 and
PKS~B1210$-$52, are part of the Galactic supernova remnant
G296.5$+$10.0 \citep{1209-52.snr}, and have not been included.  The
third, PKS~B1400$-$33, is a low-surface-brightness possible relic
radio source associated with the weak cluster around NGC\,5419
\citep{rad.G4,rad.S17}; this source was measured by SM75 and has been
included in the MS4 sample.

The Parkes surveys did not have a uniform cutoff in Galactic latitude,
and missed some regions with $|b| > 10^{\circ}$.  These excluded
regions, however, occupy only a small fraction of the sky south of
$-30^{\circ}$ (\citealp{jekers}{, Figure~1}), and given the rarity of
sources as diffuse as PKS~B1400$-$33, it is unlikely that any other
low-surface-brightness sources have been overlooked.

Completeness may be affected by the accidental omission of extended
sources with components listed separately in the MRC.  A giant source
with compact lobes separated by $>8'$ would have no extension flag in
the MRC, and would not be in our list of extended candidates.  There are
several giant radio galaxies in Table~\ref{tab2.extra}, and we estimate
that very few, if any, have been missed.

Errors in estimating $S_{\rm 408}$ from the radio spectrum may affect
completeness, particularly for values relying on extrapolation.  The
843\,MHz integrated flux densities of JM92 were later found to be
systematically low (see Section~\ref{843.posflux}) by 6\%; this could,
in principle, lead to an extrapolated $S_{\rm 408}$ for a genuine
4\,Jy source being reduced below the sample limit.  Extended sources
falling just below the sample limit are listed in
Table~\ref{tab3.nearmiss}.

\subsubsection{Random Errors}

The sample content may also be biased by random errors in the flux
density measurements, as some sources with measured $S_{\rm 408}
> 4.0$\,Jy may have true flux densities $< 4.0$\,Jy, and vice versa.
Because the differential source counts have a negative slope near 4 Jy,
slightly more sources will be wrongly included in the sample than will
be wrongly excluded from it \citep{edbias.jauncey}.  We estimated the
number of sources likely to be affected by considering the flux
densities and flux density errors of all MRC sources in the region of
interest with measured flux densities between 1 and 10\,Jy.  Assuming
the MRC flux-density errors were Gaussian, we calculated the expected
number of sources with measured $S_{\rm 408}$ between 4 and 10\,Jy
but true $S_{\rm 408} < 4.0$\,Jy to be 6.1.  Similarly we calculated
the expected number of sources with measured $S_{\rm 408}$ between
1 and 4\,Jy but true $S_{\rm 408} > 4.0$\,Jy to be 5.7.  

The net number of sources wrongly included is therefore $<1$, so we
can be confident that the sample {\it size\/} is not affected,
although the sample {\it content\/} is not quite the same as it would
be for zero measurement error.  As the MRC errors are not actually
Gaussian but have a long tail \citep{rad.L1}, the numbers of sources
thus brought into and taken out of the sample may be slightly larger
than estimated here.

\subsubsection{Variability}

Flux-density variability can also cause a bias in sample content, but
as the MS4 contains comparatively few flat-spectrum sources, only a
small fraction of the sample can be affected by variability.  While
30--50\% of flat-spectrum sources are variable at 408\,MHz
\citep{lfv.mant}, less than $2$\% of steep-spectrum sources vary at low
frequency \citep{lfv.nosteep2,lfv.nosteep1}.  As only 14 of the 228 MS4
sources have flat radio spectra ($\alpha \geq -0.5$, defined in the
sense $S_{\nu} \propto \nu^{\alpha}$), fewer than $5$\% of all MS4
sources are likely to vary significantly at 408\,MHz.  Consequently the
effects of variability on sample size have been ignored.

\subsubsection{Summary}

In summary, problems of incompleteness are likely to affect the net
sample size only for extended sources.  Fewer than five extended MRC
sources are likely to have been wrongly excluded from the MS4 sample
because of systematic underestimates of their flux densities.  From
examination of the lower resolution Parkes 408\,MHz surveys, few if any
sources have been excluded through not being in the MRC.  When the
survey with the Mauritius radio telescope \citep{maurit} at 151.5\,MHz
has been analysed, it will provide an extra point in the radio spectrum
to verify $S_{408}$ for extended sources north of $\delta =
-70^{\circ}$.

\subsection{Comparison with the Northern Sample of \protect\citet{best.samp}}
\label{sec.bestcomp}

A sample similar to the MS4 has been defined from the MRC to
have $S_{408} > 5$\,Jy in the declination range $-30^{\circ} \leq
\delta \leq +10^{\circ}$ (\citealp{best.samp}, hereafter BRL99).
Although containing useful radio and optical data, the BRL99 sample
has the disadvantage --- for statistical studies --- of radio
incompleteness, as it was defined purely from the MRC flux densities,
without examining those extended sources with MRC flux densities below
5\,Jy.

To assess the number of sources likely to be missing from the BRL99
sample, we used the flux densities of the MS4 sample.  In total, 160
MS4 sources had integrated $S_{408} > 5$\,Jy; of these 15 were
extended sources with $S_{408} < 5$\,Jy in the MRC.  The median
angular size of these 15 sources was 450$''$ and the minimum angular
size was 126$''$.  These sources form around 10\% of the MS4 sample
with $S_{408} > 5$\,Jy.  We would expect a similar fraction to be
missing from the BRL99 sample.  The incompleteness is most likely to
affect their sample at low redshifts, as they predict. We note that of
the 15 MS4 sources $\geq 5$\,Jy which would have been excluded by the
BRL99 selection criteria, 14 have $z < 0.2$.  This deficit is evident
in the lowest redshift bin of Figure 53 of BRL99, and will necessarily
compromise studies of the local radio luminosity function using this
sample.

In assessing the completeness of their sample, BRL99 have misguidedly
used the PKSCAT90 compilation \citep{pkscat90} rather than the
original Parkes catalogues \citep{rad.B1,pks.-2000,pks.0020}.  Most of
the 408\,MHz flux densities listed by PKSCAT90 are not Parkes
measurements at all, but are taken directly from the MRC.  This
explains the large number of data points in Figure~52 of BRL99 for
which {\it exactly\/} 100\% of the so-called ``Parkes flux density''
is contained in the MRC value.  Unfortunately, there is no rationale
given by the compilers of PKSCAT90 for when they chose to use MRC
rather than Parkes data.  The puzzling bimodal distribution in
Figure~52 of BRL99 for angular sizes greater than $100''$ suggests
that the decision was {\it not\/} related to angular size.  As a
result, flux densities, spectral indices, and radio powers for large
angular size sources in Table 2 of BRL99 will be unreliable.

\subsection{The MS4 Sample}
\label{sec.thesample}

The sample contains 228 sources, and is summarised in
Table~\ref{tab4.s408}.  The columns of this table are as follows:

\begin{enumerate}

\item MRC name.

\item Parkes name.  If there are two digits of declination this is from
the original Parkes catalogue \citep{rad.B1,rad.P1}, and if three
digits, from the Parkes 2700\,MHz catalogue (\citealp{2700.last} and
references therein).

\item Name of the source as given in the 85.5\,MHz MSH survey
\citep{rad.M1,rad.M2}.

\item 408\,MHz flux density.  This is usually from the MRC but for some
extended sources is taken from other 408\,MHz measurements or estimated
from the radio spectrum (see Section~\ref{sec.extflux}).  References
are given in Column~\ref{tab5.s408} of Table~\ref{tab5}.

\item References to radio images in the literature, excluding VLBI
images.

\end{enumerate}

\begin{figure}
\begin{center}
\includegraphics[clip,scale=0.63,angle=180]{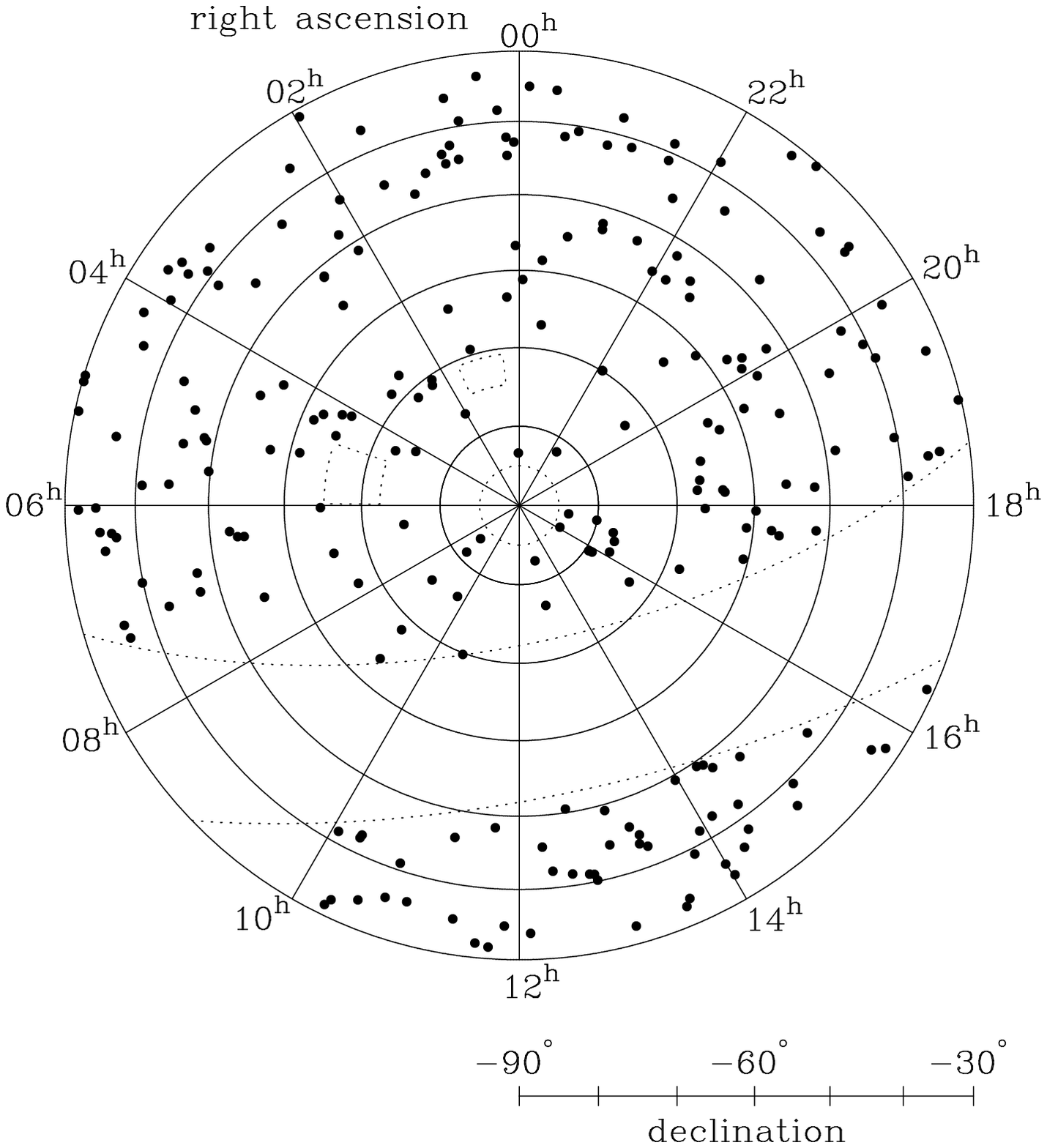}
\end{center}
\caption{Plot of the source locations on a Lambert equal-area
projection of the sky.  The South Celestial Pole is at the centre of the
plot.  Dotted lines mark the borders of regions excluded from the sample:
the Galactic plane, the circumpolar region, and the Magellanic Clouds.} 
\label{fig1.sky} 
\end{figure}

The sample covers an area of 2.43\,sr, with a source density of 94
sr$^{-1}$, compared with the 3CRR sample which covers an area of 4.24
sr and has a source density of 41 sr$^{-1}$.  The higher source
density is mainly due to the difference in flux-density cutoff. The
3CRR sample is defined by $S_{\rm 178} \geq 10.9$\,Jy, based on
the flux-density scale of \citet{baars77} which agrees (at 408 MHz)
with that of \citet{flux408.wyllie} to 3\%.  A source with $S_{\rm
178} = 10.9$\,Jy and spectral index of $\alpha = -0.81$, the
median value for the 3CRR sample, will have $S_{\rm 408} =
5.6$\,Jy.  The sky distribution of the MS4 sources is shown in
Figure~\ref{fig1.sky}.

The higher selection frequency means that the MS4 sample will have
slightly different properties from the 3CRR, containing more
flat-spectrum sources.  Because of the shape of the radio luminosity
function, the lower flux-density cutoff means that the MS4 sample will
contain a larger fraction of sources at high redshift.

\section{OBSERVATIONS AT 843\,MHZ}
\label{sec.mostobs}

The sources have all been imaged with the Molonglo Observatory
Synthesis Telescope (MOST) at 843\,MHz.  With better resolution and
sensitivity than the 408\,MHz Molonglo Cross, the MOST provides
improved information about source positions, structures, and blends.
Flux densities at 843\,MHz conveniently bridge the gap between 408 and
1400\,MHz.  The positional accuracy of about $1''$ available with MOST
makes it possible to obtain unambiguous optical identifications for
most strong sources, even when the optical counterpart is faint.

The MOST is described by \citet{most.mills81}, \citet{most.jgr91}, and
references therein.  It was constructed from the 1.6~km east-west arm of
the 408~MHz Molonglo Cross Telescope, and modified to operate at
843\,MHz with a bandwidth of 3\,MHz.  Rather than recording complex
visibilities, the MOST forms a comb of 128 fan beams in real time.  An
image of the sky is reconstructed from the fan-beam responses using 
back-projection \citep{perley.syn}.  The synthesised beam
is $43'' \times 43'' \cosec |\delta|$, and the basic field size is $23'
\times 23' \cosec |\delta|$; beam switching allows field sizes of up to
$160'$ to be observed \citep{sumss}.  It requires twelve hours for full
hour-angle coverage, in which case the $uv$ plane is totally filled
within an ellipse, apart from a small hole at the centre corresponding
to the 15\,m gap between the East and West arms.

\subsection{`CUTS' Observing}
\label{sec.cutobs}

Because of the large number of sources, the observations were done in
`CUTS' mode, a method of time-sharing in which about ten sources are
observed in one 12-hour session.  Most of the CUTS observing was
performed by R.W.H. in 1985 and 1986, as part of a program to establish a grid
of calibrators for MOST \citep{rad.C1}.  Each source was observed in
a cyclic schedule under computer control, with typical dwell times
(CUTS) of 4 minutes at each of eight widely spaced hour angles. One or
two unresolved sources were included as calibrators in each cycle.

In total 196 sources from the MS4 sample were observed in these CUTS
runs, with the remainder being covered by full synthesis observations.
Data reduction was performed using in-house software written by
C.\ R.\ Subrahmanya, J.\ E.\ Reynolds, and T.\ Ye.  The imaging process
for CUTS is essentially the same as for full-synthesis data, but the
software also allows calibration and viewing of individual CUTS,
enabling angular sizes of compact sources to be estimated more
accurately than from two-dimensional images.

\subsection{Angular Sizes and Position Angles}
\label{sec.cutprof}

The signal in each CUT consists of a one-dimensional projection of the
sky brightness distribution, convolved with the telescope's
instantaneous point-spread function.  Because they had higher angular
resolution than the final two-dimensional images, the 1-D CUTS profiles
were used to measure angular sizes and detect incipient double
structure.  The profiles were first deconvolved, using a 1-D CLEAN
algorithm with a Gaussian restoring beam of FWHM $30''$. The dirty beam
was estimated from the median of the CUTS profiles of the calibrators.    

The largest angular extent and position angle of slightly resolved
sources were determined by fitting a rectified sinewave to deconvolved
source width as a function of hour angle.  Comparison of repeat
observations, and with higher-resolution ATCA images (\papern\/), shows
that for sources with LAS $>15''$, the position angles were accurate to
about $5^{\circ}$, and the angular sizes to about $3''$.

For sources resolved into doubles we fitted a Gaussian to each
component and measured the component separation directly.  For sources
with LAS $>1'$ which were not edge-brightened doubles, we used the
two-dimensional images to measure the angular size.

\subsection{Imaging and Deconvolution}

After removing individual CUTS affected by confusion in the fan beams,
we formed raw two-dimensional images by back-projection.  These images
were then deconvolved with the standard CLEAN algorithm, using a dirty
beam formed from the 1-D calibrator template.  An example of raw and
CLEANed CUTS images is shown in \citet{posflux.molonglo}.

The dynamic range of the CLEANed images, defined as the ratio of the
peak to the rms noise, was typically about 100:1, and was limited
mainly by sidelobes from off-field sources, and by small variations in
beam shape during the observation.  For well resolved sources (LAS
$\gtrsim 3'$), the dynamic range was also limited by the success of
CLEAN in modelling extended structure.  The dynamic range, although
lower than available from a full-synthesis observation, was sufficient
to define structure on arcminute angular scales, and to provide
reliable positions.

\begin{figure}
\includegraphics[bb=50 57 585 748,clip,scale=0.82]{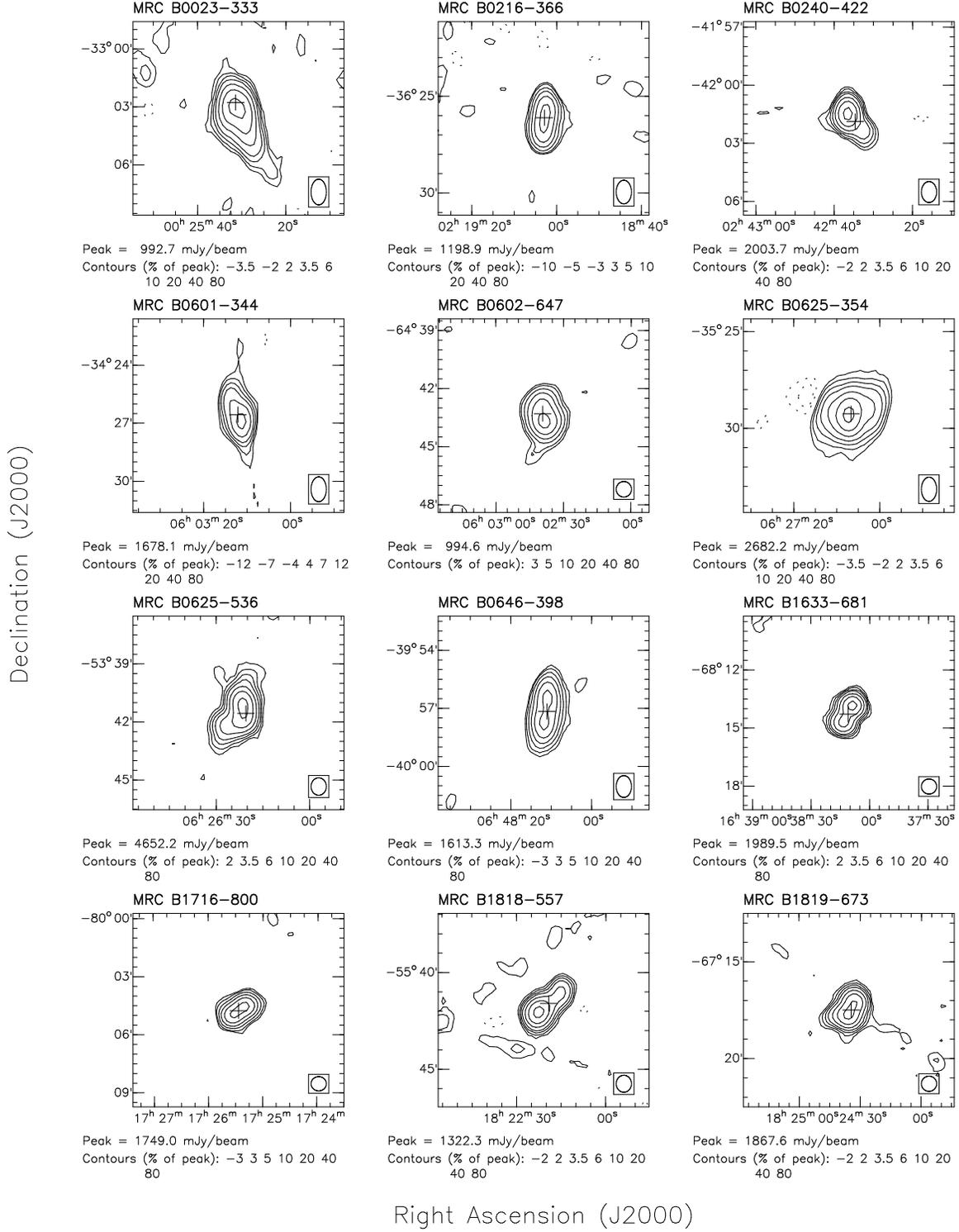}
\caption{843\,MHz contour images of slightly extended MS4 sources,
obtained in CUTS observing mode (see
\S~\protect\ref{sec.cutobs}).  The crosses mark the positions of
the optical counterparts. The Gaussian restoring beam is shown in the
bottom right-hand corner of each plot.  The side length of each plot is
10$'$.}
\label{fig2.maps} 
\end{figure}

Figure~\ref{fig2.maps} contains MOST CUTS images of a selection of 12
MS4 sources which appear extended at 843\,MHz but which were not noted
as extended in the MRC; all have angular extent $<2'$.

\subsection{Positions and Flux Densities}
\label{843.posflux}

Peak and centroid positions, and peak and integrated flux densities
were measured from the CLEANed images.  The peak positions and flux
densities were obtained using a biquadratic fit to the pixel of
highest flux density and the four pixels surrounding it.

\subsubsection{Flux Densities}

Integrated flux densities were obtained using the method of
\citet{paj.thesis}.  This involved considering the source brightness
distribution as a surface in three-dimensional space, with the
vertical axis proportional to flux density per beam area.  The volume
beneath this surface, and above a given base level, was plotted as a
function of base level and then extrapolated to a base level of zero.
A zero base level was used because there was no evidence of a negative
bowl --- the absolute value of the mean flux density in the area
surrounding the sources was well below the rms noise.

The flux-density scale used is that of \citet{mrc.thesis}, which was
based on interpolation between 408\,MHz Molonglo and 2700\,MHz Parkes
flux densities.  Comparison of this scale with that of
\citet{flux843.caganoff}, based on absolute flux-density measurements at
843\,MHz, yielded a mean flux-density ratio of $0.99 \pm 0.03$ for seven
strong sources \citep{posflux.molonglo}.  No northern hemisphere
843\,MHz measurements are available for further comparisons.  We
estimate the external accuracy of the flux-density calibration to be
about $\pm$5\%.  Based on repeated observations, we estimate the total
flux-density error for compact sources (LAS $\leq 1'$) to be $\pm$7\%
and for extended sources (LAS $> 1'$) to be about $\pm$10\%.

Comparison of integrated flux densities of 32 sources in the present
study with the values reported by JM92 showed a significant systematic
difference, with a mean ratio JM92/MS4 of $0.94\pm0.02$, with a
standard deviation of 0.12.  The difference was significant at the 2\%
level, using a two-tailed Wilcoxon signed-rank test on the fractional
difference in flux density.  While the cause of this difference is
unclear, the CUTS-based flux densities were preferred because they
were more consistent with adjacent points in the radio spectra.

\subsubsection{Positions}

Position calibration of MOST is tied to a grid of unresolved sources
(LAS $<10''$) with accurate centroid positions measured at 843 MHz
\citep{posflux.molonglo}.  The calibrator errors are estimated to be
$\sim 0.''2$ rms in right ascension and $\sim 0.''2 \cosec |\delta|$
rms in declination, with systematic errors $<0''.1$ \citep{rad.C1}.

We calculated the mean offsets in right ascension and declination for
all the calibrators in each CUTS run, and used these to correct the
measured target-source positions.  
The rms scatter was $1''.0$ in $\alpha$, and $1''.1 \cosec |\delta|$
in $\delta$.  We also tested positional accuracy by comparing
positions of 16 target sources in repeated observations.  Median
differences between pairs of positions for the 12 sources with
LAS $<1'.5$ were $0''.5$ in RA and $0''.9 \cosec |\delta|$ in Dec, and
for the four sources with LAS $>1'.5$, the median differences were
$1''.5$ in RA and $1''.9\cosec |\delta|$ in Dec.  The error for
extended sources is larger because of the uncertainties in determining
the centroid at low flux-density levels.

The adopted position errors for the CUTS observations are therefore
$\Delta\alpha = 1''.0$, $\Delta\delta = 1''.1 \cosec
|\delta|$ for sources with angular extent $<1'.5$, and $\Delta\alpha =
1''.5, \Delta\delta = 1''.9 \cosec |\delta|$ for sources with angular
extent $>1'.5$.

\section{ESTIMATION OF FLUX DENSITIES AT 178\,MHZ}
\label{sec.est178}

To compare the properties of the MS4 sources with those in the northern
3CRR sample, flux densities at 178\,MHz were estimated by interpolation
or extrapolation from the radio spectrum.  In
Section~\ref{sec.discuss}, these flux densities will be used to define
a southern 3CRR-equivalent sample.

\begin{figure}
\begin{center}
\includegraphics[bb=0 50 559 645,scale=0.7,clip=]{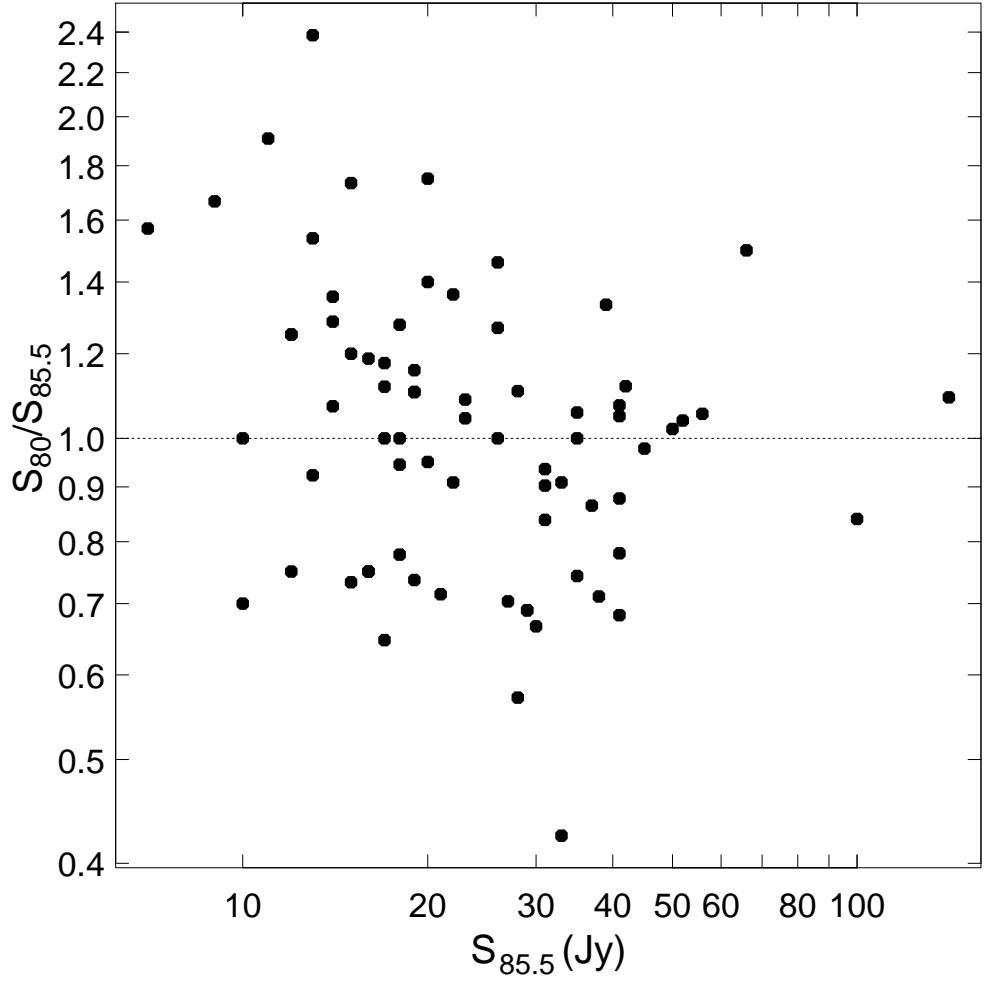}
\end{center}
\caption{Ratio of $S_{\rm 80 MHz}$ to $S_{\rm 85.5\,MHz}$ versus
$S_{\rm 85.5\,MHz}$.  The values at 80\,MHz were measured with the
Culgoora Circular Array, and those at 85.5\,MHz were from the MSH survey.}
\label{fig3.s80} 
\end{figure}

For most sources the only flux-density measurements below 178\,MHz were
those from the 85.5\,MHz survey of Mills, Slee, and Hill
\citep{rad.M1,rad.M2}, and those at 80 and 160\,MHz measured with the
Culgoora radioheliograph (\citealp{rad.S9}, \citealp{rad.S10} and
\citealp{rad.S11}; revised by \citealp{rad.S12}).  Because of their
importance for interpolation, the 85.5\,MHz MSH and 80\,MHz Culgoora
values were compared to look for any systematic differences.  The
Culgoora flux densities, measured with a $3'.7$ beam, were expected to
be low for some extended sources, while the MSH values, measured with a
50$'$ beam, were likely to be overestimated due to confusion.  It was
therefore expected that the MSH values would be higher on average.  If
anything, the plot in Figure~\ref{fig3.s80} shows that the opposite is
true, although the scatter is large.  Since the plot does not indicate
a clear preference, both 80 and 85.5\,MHz values were used in the fit
(if available), unless other information led us to prefer one of the
values.

Interpolation or extrapolation was done by performing a weighted
least-squares polynomial fit to $\log S_{\nu}$ as a function
of $\log \nu$.  A quadratic fit was used in the first instance.
If the reduced $\tilde{\chi}^2$ for the fit was greater than 5, or
if individual data points with small listed errors lay far from the
fitted curve, a linear or cubic fit was attempted, depending on the
apparent curvature of the spectrum and on the number of data points.
Anomalous data points were excluded from the fit if there was evidence of
confusion or over-resolution, or if the source was known to be variable.
As the object was to optimise the fit at 178\,MHz, high-frequency points
($\nu \geq 2700$\,MHz) were more likely to be excluded if they did not
fit well with the low frequency spectrum.

Our estimates of $S_{\rm 178}$ are listed in Column~\ref{178col}
of Table~\ref{tab5}; the order of the polynomial fit to $\log \nu$ is
given to the right of the flux density.  Comments are given below for
sources in which the fit may be unreliable, based either on a large
$\tilde{\chi}^2$ or on other information.

\subsection{Comments on Individual 178\,MHz Estimates}

\cind{MRC~B0008$-$421} Required extrapolation below 408\,MHz; estimated
$S_{\rm 178}$ is unreliable as the radio spectrum is curved.

\cind{MRC~B0240$-$422} The value of $S_{\rm 160} = 11.5$\,Jy is
higher than expected from the fit; if correct, this source should
be included in our southern equivalent of the 3CRR sample
(\S~\ref{sec.discuss}).

\cind{MRC~B0438$-$436} Very compact source; the fit is probably affected by
flux-density variability.

\cind{MRC~B0454$-$463} Poor fit at low frequency ($\tilde{\chi}^2 =
11.5$).  A 5\,GHz ATCA image (Paper\,2) shows the radio structure to
be core-dominated; the source is known to vary at high frequency
\citep{rad.W1,rad.W3}.

\cind{MRC~B0511$-$484} Most flux densities are affected by blending with two
fainter sources; $S_{\rm 178}$ relies on extrapolation below
408\,MHz.

\cind{MRC~B0647$-$475} Compact source with a curved spectrum and flux
densities at only four frequencies.

\cind{MRC~B1315$-$460} Compact source with a curved spectrum and flux
densities at only four frequencies.

\cind{PKS~B1318$-$434} Flux densities are affected by confusion from
Centaurus~A, making the extrapolation to 178\,MHz unreliable.  A
quadratic fit to the four data points has been adopted; if the spectrum
is really straight the source is probably stronger than the 3CRR cutoff
of $S_{\rm 178} = 10.9$\,Jy.

\cind{MRC~B1445$-$468} The measurements at 80 and 85.5\,MHz are highly
discrepant, so the fitted $S_{\rm 178}$ is unreliable.

\cind{MRC~B1545$-$321} Giant radio galaxy (LAS = $7'$).  The estimate of
$S_{\rm 178}$ is unreliable as it relies on extrapolation from
843 MHz.

\cind{MRC~B1549$-$790} Flat-spectrum source, known to vary at 843 MHz
\citep{var843}; $S_{\rm 178}$ relies on extrapolation and may be
unreliable.

\cind{MRC~B1933$-$587} Poor fit to a polynomial at low frequency.  A
5\,GHz ATCA image (Paper\,2) shows the radio structure to be
core-dominated; the source is known to vary at high frequency
\citep{rad.W1,rad.W3}, but is non-variable at 843 MHz
\citep{var843}.  

\cind{MRC~B1934$-$638} $S_{\rm 178}$ for this GPS source is very
uncertain.  We used the cubic polynomial fitted by
\citet{reynolds.scale}.

\cind{MRC~B2153$-$699} Extrapolation unreliable because of blending
with MRC~B2152$-$699.

\section{SUMMARY OF THE MS4 SAMPLE}

A summary of the 843\,MHz data for the MS4 sample, along with flux
density measurements from the literature at other frequencies, is given
in Table~\ref{tab5}.  Although different measurements in the literature
have used (nominally) different flux-density scales, no correction has
been made; the figures in the table are quoted {\it directly\/} from
the papers cited.  The columns of Table~\ref{tab5} are as follows:

\begin{enumerate}

\item Source name, taken from the MRC, unless the source has more than
one MRC entry, in which case the name is from the original Parkes
catalogue \citep{rad.B1,rad.P1} or Parkes 2700\,MHz catalogue
(\citealp{2700.last} and references therein).

\item \label{tab5.ra} Right ascension in J2000 coordinates at 843\,MHz.

\item \label{tab5.dec} Declination in J2000 coordinates at 843\,MHz.
The reference for the radio position in columns~\ref{tab5.ra} and
\ref{tab5.dec} is the same as for the 843\,MHz flux density in
column~\ref{tab5.843}.

\item \label{178col} Estimated flux density at 178\,MHz, as described
in Section~\ref{sec.est178}, and the order of the polynomial used to
calculate the estimate.

\item Flux density at 80\,MHz; the observing frequency is 85.5\,MHz,
if taken from MSH, or 80\,MHz, if measured with the Culgoora Circular
Array.  The latter flux densities are from the compilation of
\citet{rad.S12}, with revised calibration.

\item \label{tab5.s408} Flux density at 408\,MHz, taken from the
MRC, except for some extended sources.  In general, flux densities
were taken from published data in the following order of preference:
SM75, MC4, \citet{rad.W6}, the Parkes catalogues, \citet{rad.E1}.  Flux
densities for a few sources were estimated using the Molonglo Transit
Catalogue or from the radio spectrum.

\item \label{tab5.843} Flux density at 843\,MHz, measured with MOST.
The values of \citet{rad.C1} were used for MOST calibrators, as were
those of \citet{rad.S16} for six giant radio sources.  For the
remaining sources the flux density was taken either from JM92 or from
the present study, depending on which image had the higher dynamic
range.

\item Flux density at 1400\,MHz.  Data are mostly from Parkes
(1410\,MHz: \citealt{rad.B1,rad.P1,rad.W2}), from Owens Valley
(1425\,MHz: \citealt{rad.F1,rad.F2}), or from the NRAO VLA Sky Survey
catalog (NVSS; 1400\,MHz: \citealt{rad.C6}).  For a few sources data
were taken from the Instituto Argentino de Radioastronomia (IAR) 30\,m
telescope (1410\,MHz: \citealt{rad.Q1,rad.Q2,rad.Q3}), from the
Fleurs Synthesis Telescope, or the Australia Telescope Compact Array.
The flux-density scale for the Parkes catalogues \citep{rad.B1,rad.P1}
and Owens Valley data was that of \citet{flux408.ckl}, whereas the data
of \citet{rad.W2} and the IAR 30\,m data were on the scale of
\citet{wills.scale}.  At 1410\,MHz the scale of \citet{wills.scale} is
1.08 times the scale of \citet{flux408.ckl} \citep{baars77}.  NVSS used
the flux density scale of \citet{baars77}.

\item Flux density at 2700\,MHz, measured with the Parkes telescope,
except for one source which was observed at 2640\,MHz with Owens Valley
\citep{rad.R1}.  2650\,MHz flux densities are from the original Parkes
catalogue or from \citet{rad.W2}, while 2700\,MHz values are from the
Parkes 2700\,MHz survey or from \citet{rad.W2}.  The Parkes 2700\,MHz
survey used a scale in which the peak flux density of Hydra~A was
23.5\,Jy; \citet{rad.W2} used the flux-density scale of
\citet{wills.scale}.

\item Flux density at 5000\,MHz, measured with the Parkes telescope.
Values at 4850\,MHz are from the Parkes-MIT-NRAO survey (PMN:
\citealt{rad.G1,rad.W7}), those at 5000\,MHz are from \citet{rad.W2},
and those at 5009\,MHz are from \citet{rad.W2}, or the Parkes 2700\,MHz
survey, or \citet{rad.S1}, or \citet{rad.S2}.  The PMN survey used the
flux density scale of \citet{baars77}; \citet{rad.W2} used the scale of
\citet{wills.scale}; the Parkes 2700\,MHz surveys used a 5 GHz scale in
which the peak flux density of Hydra~A was $S_{\rm 5009} = 13.05$\,Jy,
as did \citet{rad.S1} and \citet{rad.S2}.

\item Radio spectral index $\alpha$ measured between 408 and 2700\,MHz,
defined in the sense $S_{\nu} \propto \nu^{\alpha}$. For sources
without measurements at both those frequencies, the spectral index was
measured either between 408 and 5000\,MHz or between 843 and
5000\,MHz.

\item \label{tab5.las} Largest angular size (LAS) at 843\,MHz.  The
reference is the same as for the 843\,MHz flux density, except for a
few sources, noted in Section~\ref{sec.com843}.  We have imaged all
sources with ${\rm LAS}<35''$ with the ATCA at 5\,GHz (\papern\/), with
the exception of MRC~B0521$-$365, MRC~B0743$-$673, MRC~B1740$-$517,
MRC~B1814$-$519, \\
MRC~B2153$-$699, and MRC~B2259$-$375.

\item Position angle of extension at 843\,MHz, defined in degrees east
of north (modulo 90).  The reference is the same as for the 843\,MHz
flux density, except for a few sources which are noted in
Section~\ref{sec.com843}.

\end{enumerate}

\subsection{Comments on Individual Sources}
\label{sec.com843}

\cind{MRC~B0208$-$512} Detected with EGRET \citep{egret.cat2}.

\cind{MRC~B0214$-$480} The Parkes 408\,MHz and MSH 85.5\,MHz flux densities
are probably affected by blending with MRC~B0211$-$479.
The position, angular size, and position angle in Table~\ref{tab5} are
from JM92, but $S_{843}$ was measured from MOST images from the SUMSS survey
\citep{sumss2003}.

\cind{MRC~B0252$-$712} The spectrum turns over at low frequency. The flux
density of $S_{\rm 85.5} = 7$\,Jy is noted as uncertain in MSH,
but corresponds well with the slight spectral curvature evident at
higher frequencies.

\cind{PKS~B0319$-$45} Giant radio galaxy.  Because of the large angular
size, the flux densities are uncertain.  The MSH 85.5\,MHz measurement
was affected by sidelobes of Fornax~A.

\cind{MRC~B0320$-$373} Fornax~A.

\cind{MRC~B0336$-$355} Blended at low resolution with the foreground
source MRC~B0336$-$356.  At 2650 and 5009\,MHz, the peak flux densities
\citep{rad.W2} were used, as they are expected to be less affected by
blending.

\cind{MRC~B0411$-$346A} The weaker source MRC~B0411$-$346B is unrelated.

\cind{MRC~B0453$-$301, MRC~B0456$-$301} MSH\,04$-$3{\it 14} is a blend
of these two sources.

\cind{MRC~B0506$-$612} The 8.87 GHz flux density of 2.63\,Jy
\citep{rad.S5} is substantially higher than expected from the radio
spectrum, probably indicating high-frequency variability.

\cind{MRC~B0511$-$484} Unequal double, with two neighbouring weaker sources,
presumed to be unrelated.  The SM75 integrated value of $S_{\rm
408} = 8.8$\,Jy is affected by blending.  The MRC fitted value of
6.84\,Jy was preferred but is still uncertain.  The Parkes flux
densities at 2.7 and 5 GHz are also affected by blending, so the radio
spectrum is not well determined.

\cind{MRC~B0518$-$458} Pictor~A.  

\cind{MRC~B0521$-$365} Well studied blazar, detected as a gamma-ray source
by EGRET \citep{egret.cat2}.

\cind{PKS~B0707$-$35} The flux density $S_{\rm 5000} = 0.84$\,Jy
\citep{rad.W5} appears anomalously high. 

\cind{MRC~B0743$-$673} The 1410 MHz flux density \citep{rad.P1} is higher
than expected from the radio spectrum, and probably affected by blending
with MRC\,B0742$-$674.

\cind{MRC~B1017$-$421, MRC~B1017$-$426} MSH\,10$-$4{\it 4} is a blend
of these two sources.

\cind{MRC~B1136$-$320} In the Texas catalog \citep{rad.D2} the double
separation is quoted as 58$''$ in P.A. $-15^{\circ}$, consistent with
the values in Table~\ref{tab5}.

\cind{MRC~B1143$-$316} The Texas catalog \citep{rad.D2} lists this
source as a double with separation $34''$ in P.A. $86^{\circ}$.  The
MRC value of $S_{\rm 408} = 5.77$\,Jy is preferred to SM75's
anomalously low value of 3.8\,Jy.

\cind{MRC~B1302$-$491} Nearby edge-on spiral galaxy NGC\,4945, identified
by \citet{rad.M1}.  

\cind{PKS~B1318$-$434} Complex double source, with bent edge-darkened
lobes, lying behind the southern lobe of Centaurus~A
\citep{rad.C4,map.cena.haynes}.

\cind{MRC~B1322$-$427} The well studied low-luminosity radio galaxy
Centaurus~A (see review by \citealp{cena.review}).  The flux density at
843 MHz was determined from the radio spectrum.

\cind{PKS~B1400$-$33} The only source in the sample which is not in the
MRC.  An unusual extended source of low surface brightness.  There is a
nearby compact source associated with the E0 galaxy NGC\,5419 (see
\citealp{rad.E1}).  \citet{rad.G4} suggest that the extended component
may be a relic radio source associated with the poor cluster Abell S753
around NGC\,5419.  The radio properties have been studied in detail by
\citet{rad.S17}; we have used their flux densities at 843 and 1398\,MHz.

\cind{MRC~B1425$-$479} Because of the angular size of $4'.5$, the values
of $S_{\rm 408}$ and $S_{\rm 5000}$ in Table~\ref{tab5} may be slightly
underestimated.

\cind{MRC~B1445$-$468} The published values of $S_{\rm 80}$ --- 2\,Jy
\citep{rad.S12} and 13\,Jy \citep{rad.S9} --- are grossly
discrepant. We prefer the latter value, after multiplying it by 1.1 to
correct the flux-density scale \citep{rad.S11}.

\cind{MRC~B1549$-$790} Flat-spectrum source.  The flux density at
1410\,MHz \citep{rad.P1} is affected by blending with MRC~B1547$-$795.

\cind{MRC~B1814$-$637, MRC~B1817$-$640} MSH\,18$-$6{\it 1} is a blend
of these two sources.

\cind{MRC~B1917$-$546} Ultra-steep spectrum source.  The 1410 MHz flux
density may be affected by blending with two weak neighbouring sources
(Hunstead 1972).

\cind{MRC~B1934$-$638} Archetypal Gigahertz-Peaked-Spectrum (GPS)
source \citep{disc.1934}.

\cind{MRC~B1940$-$406} Blending with three weaker sources (JM92) 
makes the radio spectrum uncertain.

\cind{MRC~B2006$-$566} Very extended, diffuse source, associated with the
cluster Abell\,3667 \citep{aco}.  The flux density of $S_{\rm 843}
= (5.5 \pm 0.5)$\,Jy \citep{rad.R2} was obtained from a MOST
full-synthesis image after subtracting the contribution from background
sources within the source envelope (0.6\,Jy); the discrepancy with JM92
is probably due to the higher noise level in their image.  

\cind{MRC~B2052$-$474} 
Detected with EGRET \citep{egret.cat2}.

\cind{MRC~B2122$-$555} The MRC position appears to be in error, lying
$30''$ east of the MOST position.  The MRC flux density of 4.05\,Jy is
also lower than expected from the radio spectrum, suggesting that the
observation may have been affected by a large ionospheric wedge
(Hunstead 1972).  Both Parkes positions \citep{rad.W1,rad.G1} agree
with the MOST centroid.

\cind{MRC~B2152$-$699} Some flux densities are affected by blending
with MRC~B2153$-$699.  The 1415\,MHz measurement \citep{rad.C5} was
made with the Fleurs Synthesis Telescope.  At 2650\,MHz, the peak flux
density \citep{rad.W2} was used, to reduce any contribution from
MRC~B2153$-$699.

\cind{MRC~B2153$-$699} Double, extended along a position angle similar
to that of MRC~B2152$-$699.  Flux densities are affected by blending
with MRC~B2152$-$699: both the MRC value of $20.9 \pm 1.4$\,Jy and the
SM75 value of $6.0 \pm 0.4$\,Jy are regarded as unreliable.  Therefore,
the flux densities at 843 and 1415\,MHz, together with the 468 MHz
value of $10.7 \pm 0.9$\,Jy \citep{rad.E2}, have been used to estimate
the flux density at 408 MHz.

\section{DISCUSSION}
\label{sec.discuss}

The 178\,MHz flux densities estimated in Section~\ref{sec.est178} have
been used to define a strong source subsample of MS4 which we call
SMS4.  It has been chosen to have the same flux-density cutoff,
10.9\,Jy, as the northern 3CRR sample, and contains 137 sources,
compared with 172 in 3CRR.

\begin{figure}
\begin{center}
\includegraphics[bb=126 31 462 713,clip=,scale=0.8]{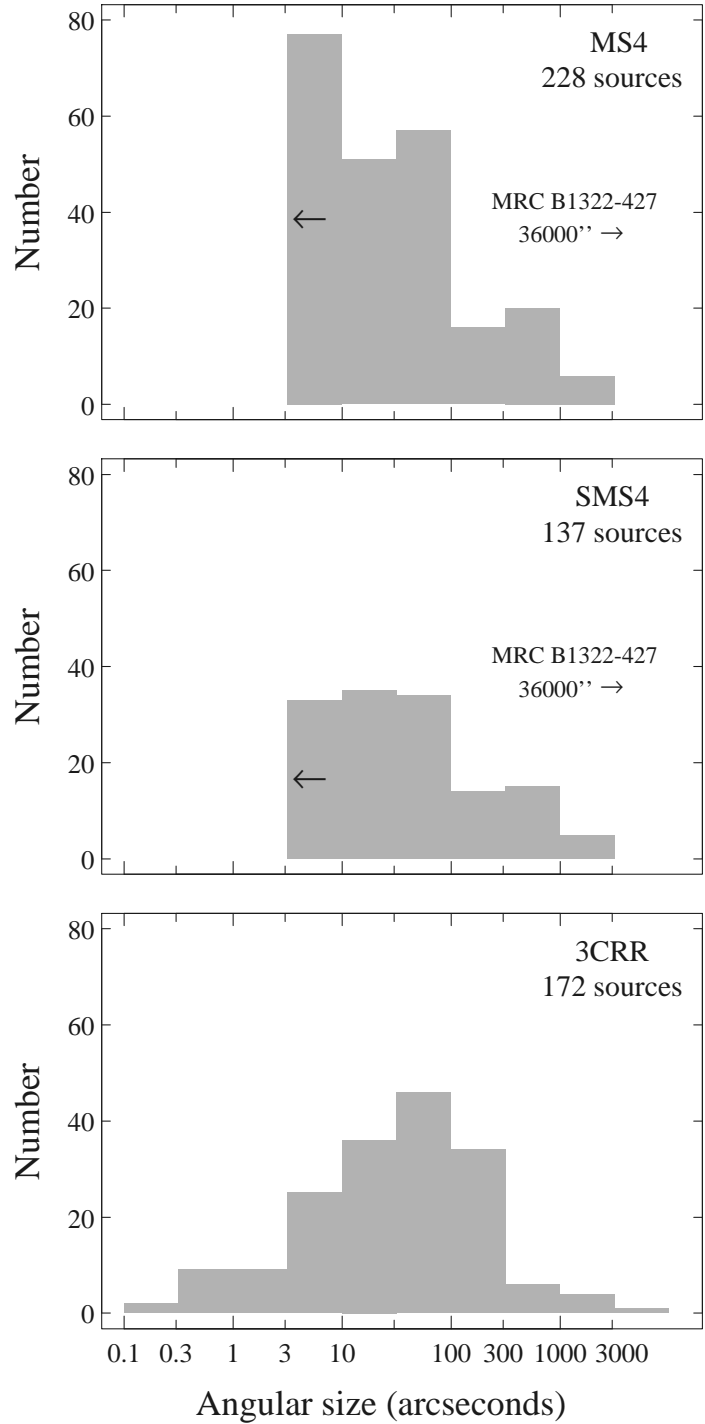}
\end{center}
\caption{Histograms of angular size for the MS4, SMS4, and 3CRR samples.
In the MS4 and SMS4 data, MRC~B1322$-$427 is off-scale, and the
leftmost bin represents an upper limit to the angular size for
many of the sources.} 
\label{fig4.las} 
\end{figure}

Comparison of SMS4 with 3CRR (Table~\ref{tab6.medians}) shows the
southern sample to have a slightly higher source density, but only at
the $2.8\sigma$ level of significance.  The difference should be
treated with caution and may simply reflect biases in the way each
sample was compiled. We have identified three possible causes of such
a bias:

\begin{itemize}
\item[(i)] As the spectra of many radio sources turn over at low
radio frequency, extrapolation from high frequencies is more likely
to overestimate than underestimate $S_{\rm 178}$.  The survey
with the Mauritius radio telescope \citep{maurit} at 151.5\,MHz will
be valuable for checking flux densities of sources north of $\delta =
-70^{\circ}$, and testing for such a bias in the SMS4 sample.

\item[(ii)] Because of the steep slope of the radio source counts, a
small systematic difference in flux-density scale can strongly bias the
source density.  The Mauritius values will be useful for checking such
effects in SMS4.

\item[(iii)] The 3CRR may be missing sources of low surface
brightness.  The angular size distributions are compared in
Figure~\ref{fig4.las}.  The median angular size of SMS4 is 32$^{+6}_{-5}$
arcsec, compared with 35.5$^{+8.7}_{-7.5}$ arcsec for 3CRR.  Although
the medians are similar, the distributions appear different:
the SMS4 has proportionally more sources with angular size 
$>300''$, and proportionally fewer sources with angular size
between 100$''$ and 300$''$.  However, because of the small numbers of
sources involved, it is not possible to draw firm conclusions about
differences in the angular size distributions.

\end{itemize}

\begin{figure}
\begin{center}
\includegraphics[bb=132 71 466 735,clip=,scale=0.9]{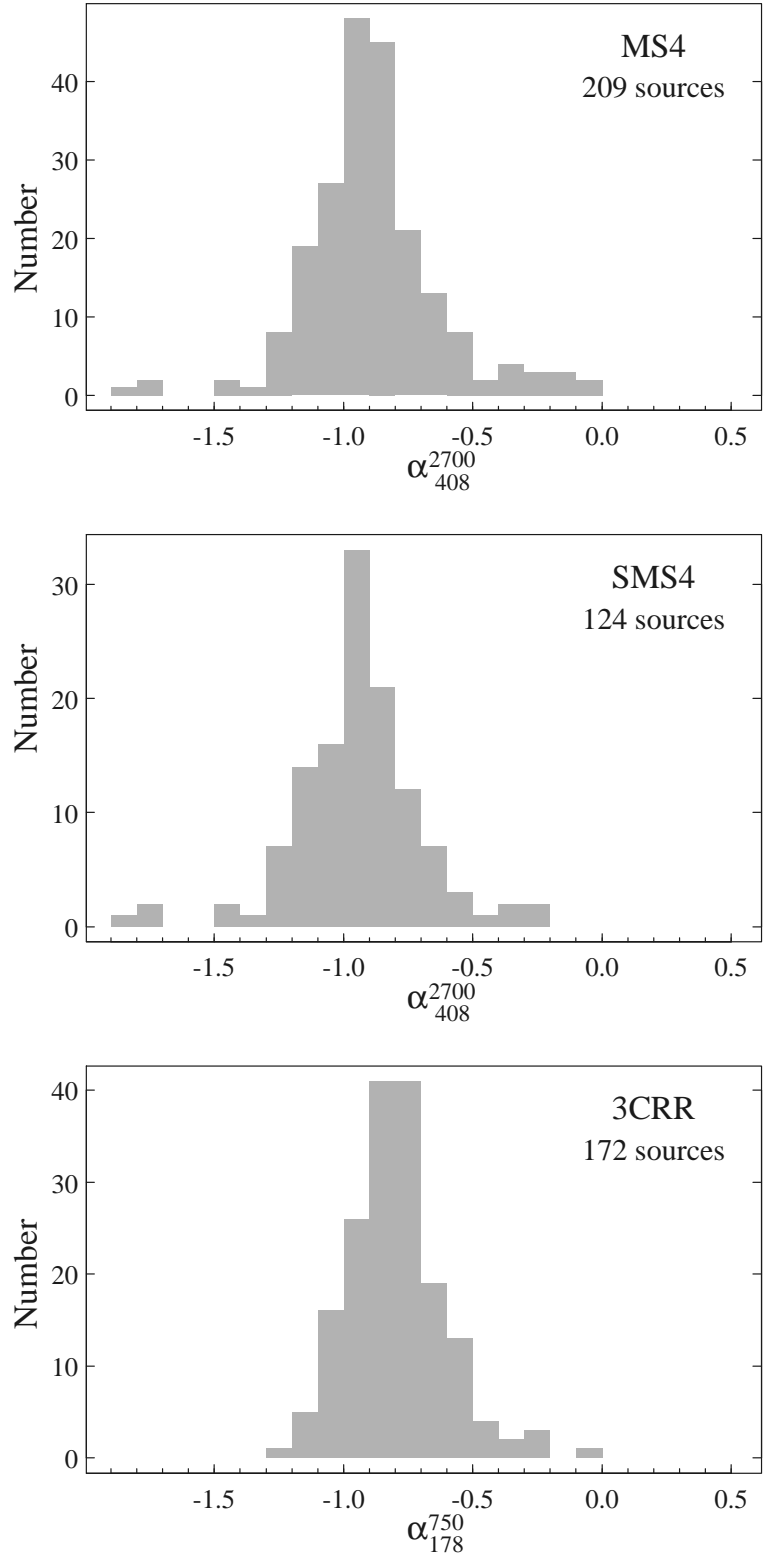}
\end{center}
\caption{Histograms of spectral index for the MS4, SMS4, and 3CRR samples.
MS4 and SMS4 sources without reliable flux density data have been omitted.}
\label{fig5.alpha} 
\end{figure}

The median flux density of SMS4 ($S_{\rm 178} = 15.7$\,Jy), is
consistent with the median value of 15.6\,Jy for 3CRR.  This is to be
expected if the radio source counts are similar.  Histograms of
spectral indices for the MS4, SMS4, and 3CRR samples are plotted in
Figure~\ref{fig5.alpha}.  The distributions for the MS4 sample show a
longer tail towards flatter spectral indices than do those of the SMS4
or 3CRR.  This is to be expected given the higher selection frequency
of the MS4, as flat-spectrum sources will be included which would be
missed with a higher flux-density cutoff and lower selection
frequency.  The median spectral indices, $\alpha=-0.91$ for SMS4 and
$-0.81$ for 3CRR (Table~\ref{tab6.medians}), are not consistent, but
can be explained simply by the different frequency ranges used to
measure $\alpha$ for each sample: 408--2700\,MHz for SMS4 and
178--750\,MHz for 3CRR.  Given that the spectra of many radio sources
turn over at low frequency, and that radio spectra often steepen at
high frequency, the flatter median spectral index of the 3CRR sources
can be explained by the lower frequency range over which $\alpha$ was
measured.  A comparison over more similar frequency ranges will be
possible when 151.5\,MHz data from Mauritius become available.

\section{SUMMARY}

The Molonglo Southern 4\,Jy (MS4) sample, a complete sample of radio
sources south of $\delta=-30^{\circ}$, has been selected from the
Molonglo Reference Catalogue (MRC) to have $S_{\rm 408} > 4.0$\,Jy.
All the MS4 sources have been imaged with MOST at 843\,MHz, thus
providing more accurate positions than available from the MRC.  The
positional accuracy is about $\Delta\alpha = 1''.0$, $\Delta\delta =
1''.1 \cosec |\delta|$ for compact sources (angular size $<1'.5$), and
$\Delta\alpha = 1''.5, \Delta\delta = 1''.9 \cosec |\delta|$ for
extended sources (angular size $>1'.5$).  Integrated flux densities at
843\,MHz have been measured from the images, and have an accuracy of
$\sim$ 7--10\%.  Reliable angular sizes and position angles have been
measured for sources with angular extent $\gtrsim 15''$.  The median
angular size for the MS4 sample as measured with MOST is $23'' \pm
4''$.

The set of MOST images provides a useful database for optical
identification, as well as for planning higher-resolution radio
imaging, to be presented in \papern\/.  Radio flux densities at other
frequencies have been obtained from the literature, and used to
retrospectively define a stronger subsample, the SMS4, with the same
flux-density limit as the 178\,MHz 3CRR sample.  Preliminary
comparisons of SMS4 and 3CRR show that their overall properties are
similar.

\acknowledgments

\section*{ACKNOWLEDGEMENTS}

We thank staff and students of the School of Physics, University of
Sydney, for their generous help with observing and data reduction.
Special thanks to the staff of the Molonglo Telescope, for operating and
maintaining it for the demanding CUTS runs, and to T. Ye and C. R.
Subrahmanya for writing and modifying data-reduction software.

We have made use of NASA's Astrophysics Data System Abstract Service,
and the NASA/IPAC Extragalactic Database (NED) which is operated by
the Jet Propulsion Laboratory, Caltech, under contract with the US
National Aeronautics and Space Administration.  The Molonglo Observatory
Synthesis Telescope is funded by both the Australian Research Council
and the Science Foundation for Physics within the University of Sydney.
AMB acknowledges the receipt of an Australian Postgraduate Research
Award over the period of this research.

\clearpage

\clearpage

\clearpage

\clearpage

\clearpage

\begin{deluxetable}{cllrl}
\tabletypesize{\footnotesize}
\tablecaption{Radio surveys of strong southern extragalactic
sources\label{tab1.surveys}}
\tablehead{
\colhead{(1)} & \colhead{(2)} & \colhead{(3)} & \colhead{(4)} &
\colhead{(5)} \\
\colhead{Survey} & \colhead{Frequency} & \colhead{Flux density} &
\colhead{No. of} &
\colhead{References} \\
\colhead{} & \colhead{(MHz)} & \colhead{limit (Jy)} & \colhead{sources} &
\colhead{}}
\startdata
MSH & \phm{0000}85.5 & \phn\phn\phn10\phm{.00} & 2270 & 1 2 3  \\
PKS & \phn\phn\phn408\tablenotemark{a} &
  \phm{0000}4\tablenotemark{b}\phd\phn & 2133 & 4 5 6 7 8 \\
PKS 2700 & \phn\phn2700\phd\phn &
  \phn\phn$\sim$0.25\rlap{\tablenotemark{c}} &
  $\sim$6000\rlap{\tablenotemark{d}} & 9 \\
MRC & \phn\phn\phn408\phd\phn &
  \phm{0000}1.0\rlap{\tablenotemark{e}}\phn & 12141 & 10 \\
PMN & \phn\phn4850\phd\phn & \phm{0000}0.04\rlap{\tablenotemark{f}}
& 50814 &  11 12 13 14 \\
\enddata
\tablenotetext{a}{\protect\cite{rad.P1} made part of their
survey at 1410\,MHz; the survey of \protect\cite{pks.2027} was made at
635\,MHz.}
\tablenotetext{b}{The flux-density limit at 408\,MHz varied from 3 to
4\,Jy; some papers contained additional sources down to
$S_{408}=1.0$\,Jy.}
\tablenotetext{c}{The flux-density limit varies from 0.1 to 0.6\,Jy.}
\tablenotetext{d}{The total number of entries in parts 1 to 14 of the
catalogue is 6842, but some sources were observed more than once.}
\tablenotetext{e}{Complete to 1\,Jy but contains sources
down to 0.7\,Jy.}
\tablenotetext{f}{The flux-density limit varies from 0.020 to
0.072\,Jy with declination.}
\tablerefs{
(1) \citet{msh1},
(2) \citet{rad.M1},
(3) \citet{rad.M2},
(4) \citet{rad.B1},
(5) \citet{rad.P1},
(6) \citet{pks.0020},
(7) \citet{pks.-2000},
(8) \citet{pks.2027},
(9) \citet{2700.last} and references therein,
(10) \citet{rad.L1},
(11) \citet{pmn.tropical},
(12) \citet{pmn.southern},
(13) \citet{pmn.equatorial},
(14) \citet{rad.W7}.}
\end{deluxetable}

\clearpage
\begin{deluxetable}{c@{ }lp{45mm}rrl}
\tablecaption{Extended sources with MRC $S_{408} < 4.0$\,Jy which
have been added to the sample\label{tab2.extra}}
\tablehead{\multicolumn{2}{c}{(1)} & \colhead{(2)} & \colhead{(3)} &
\colhead{(4)} & \colhead{(5)} \\
\multicolumn{2}{c}{IAU} & \colhead{MRC name(s)} & \colhead{LAS} &
\colhead{Integrated} & \colhead{Source of estimate} \\
\multicolumn{2}{c}{name} & \colhead{} & \colhead{(arcmin)} &
\colhead{$S_{408}$ (Jy)} & \colhead{}}
\startdata
PKS & B0114$-$47 & 0113$-$475, 0114$-$476 & 9.5 & 6.6\tablenotemark{a} &
  SM75 \\
MRC & B0214$-$480 & 0214$-$480 & 7.1 & 5.5\tablenotemark{a} & SM75 \\
PKS & B0319$-$45 & 0319$-$453, 0317$-$456 & 25.6 & 8.3 & Radio spectrum \\
PKS & B0332$-$39 & 0332$-$391A, 0332$-$391B & 7\phd\phn & 4.2 & SM75 \\
MRC & B0424$-$728 & 0424$-$728 & 3.6 & 4.3 & MC4 \\
MRC & B0429$-$616 & 0429$-$616 & 1.9 & 4.4\tablenotemark{a} & MC4 \\
PKS & B0511$-$30 & 0511$-$305, 0511$-$304     & 10.6 & 7.8 & SM75 \\
PKS & B0707$-$35 & 0707$-$359A, 0707$-$359B   & 8.1 & 4.6 & SM75 \\
\enddata
\tablenotetext{a}{PKS 408\,MHz flux density was preferred.}
\end{deluxetable}

\clearpage
\begin{deluxetable}{lrrl}
\tablecaption{Extended sources found to be just below the 4.0\,Jy
flux-density limit\label{tab3.nearmiss}}
\tablehead{
\colhead{(1)} & 
\colhead{(2)} & 
\colhead{(3)} & 
\colhead{(4)} \\
\colhead{Source} & 
\colhead{LAS} & 
\colhead{$S_{408}$} & 
\colhead{Means of} \\
\colhead{name} & 
\colhead{(arcmin)} & 
\colhead{(Jy)} & 
\colhead{estimate}}
\startdata
MRC B0816$-$705 & 6\phd\phn & 3.5 & MC4 \\
MRC B1011$-$316 &  1.3 & 3.9 & Transit catalogue \\
MRC B1148$-$353 &  1.8 & 3.8 & Transit catalogue \\
MRC B1308$-$441 & 14.4 & $\sim$3.8 & Radio spectrum \\
MRC B1511$-$315 &  2.1 & 3.8 & Radio spectrum \\
MRC B1821$-$583 &  3.3 & 3.3 & SM75 \\
\enddata
\end{deluxetable}

\clearpage 
\begin{deluxetable}{lrllr@{ }c@{ }rl}
\tabletypesize{\scriptsize}
\tablewidth{320pt}
\tablecolumns{8}
\tablecaption{List of sources in the MS4 sample.\label{tab4.s408}}
\tablehead{
\colhead{(1)} & & \colhead{(2)} & \colhead{(3)} &
\multicolumn{3}{c}{(4)} & \colhead{(5)} \\
\colhead{MRC} & & \colhead{PKS} & \colhead{MSH} &
\multicolumn{3}{c}{$S_{408}$} & \colhead{References for} \\
\colhead{name(s)} & & \colhead{name(s)} & \colhead{name} &
\multicolumn{3}{c}{(Jy)} & \colhead{radio images}}
\startdata
B0003$-$567 & & B0003$-$56  & 00$-$5{\it 1}  & 5.19 & $\pm$ & 0.23 & ... \\
B0003$-$428 & & B0003$-$42  & 00$-$4{\it 2}  & 5.15 & $\pm$ & 0.16 & ... \\
B0003$-$833 & & B0004$-$83  & \nodata & 6.62 & $\pm$ & 0.31 & 1 \\
B0007$-$446 & & B0007$-$44  & 00$-$4{\it 3}  & 6.81 & $\pm$ & 0.21 & ... \\
B0008$-$421 & & B0008$-$42  & \nodata & 6.61 & $\pm$ & 0.21 & ... \\
B0012$-$383 & & B0012$-$38  & 00$-$3{\it 5}  & 4.87 & $\pm$ & 0.14 & ... \\
B0013$-$634 & & B0013$-$63  & 00$-$6{\it 1}  & 6.90 & $\pm$ & 0.17 & ... \\
B0023$-$333 & & B0023$-$33  & 00$-$3{\it 8}  & 4.09 & $\pm$ & 0.18 & 2 \\
\enddata
\end{deluxetable}

\clearpage 
\begin{deluxetable}{l@{ }l@{\hspace{2.5mm}}l@{\hspace{2.5mm}}l@{\hspace{2.5mm}}r@{ }r@{\hspace{2.5mm}}r@{ }r@{\hspace{2.5mm}}r@{ }r@{\hspace{2.5mm}}r@{ }r@{\hspace{2.5mm}}r@{ }r@{\hspace{2.5mm}}r@{ }r@{\hspace{2.5mm}}r@{ }r@{\hspace{2.5mm}}r@{\hspace{1.5mm}}r@{ }c@{\hspace{1mm}}r}
\tablecolumns{22}
\tabletypesize{\scriptsize}
\rotate
\tablewidth{580pt}
\tablecaption{Radio positions, flux densities, angular sizes, and position
angles for the MS4 sample.  \label{tab5}}
\tablehead{
\multicolumn{2}{c}{(1)} &
\colhead{(2)} &
\colhead{(3)} &
\multicolumn{2}{c}{(4)} &
\multicolumn{2}{c}{(5)} &
\multicolumn{2}{c}{(6)} &
\multicolumn{2}{c}{(7)} &
\multicolumn{2}{c}{(8)} &
\multicolumn{2}{c}{(9)} &
\multicolumn{2}{c}{(10)} &
\colhead{(11)} &
\multicolumn{2}{c}{(12)} &
\colhead{(13)} \\
\multicolumn{2}{c}{Source} &
\multicolumn{2}{c}{RA~~(J2000)~~Dec} &
\multicolumn{14}{c}{\hrulefill~Flux density~(Jy) \hrulefill} &
\colhead{$\alpha_{408}^{2700}$} &
\multicolumn{2}{c}{LAS\tablenotemark{a}} &
\colhead{P.A.} \\
\multicolumn{2}{c}{name} &
\multicolumn{1}{l}{\llap{h}~~m~~~s} &
\multicolumn{1}{l}{~$^\circ$~~$'$~~~$''$} &
\multicolumn{2}{c}{$S_{178}$} &
\multicolumn{2}{c}{$S_{80}$} &
\multicolumn{2}{c}{$S_{408}$} &
\multicolumn{2}{c}{$S_{843}$} &
\multicolumn{2}{c}{$S_{1400}$} &
\multicolumn{2}{c}{$S_{2700}$} &
\multicolumn{2}{c}{$S_{5000}$} &
\colhead{} &
\multicolumn{2}{c}{$''$} &
\colhead{$^\circ$}
}
\startdata
MRC & B0003$-$567 & 00 05 57.64 & $-$56 28 30.6 & 8.6\phn & 2 & \phn\phn12 & [2] & 5.19 & [3] & 2.88 & [4] & 2.0\phn & [5] & 1.08 & [6] & 0.69 & [7] & $-$0.83 & 19 & m & $-$72 \\
MRC & B0003$-$428 & 00 06 01.96 & $-$42 34 41.9 & 9.5\phn & 2 & 17 & [8] & 5.15 & [3] & 2.97 & [4] & 1.7\phn & [5] & 1.02 & [9] & 0.53 & [7] & $-$0.86 & $\leq$10 & m & \nodata \\
MRC & B0003$-$833 & 00 06 14.6 & $-$83 05 60 & 13.2: & 1 & \nodata & \nodata & 6.62 & [3] & 3.40 & [10] & 2.5\phn\tablenotemark{i} & [11] & 1.19 & [12] & 0.84 & [7] & $-$0.91 & 60 & d & $-$24 \\
MRC & B0007$-$446 & 00 10 30.50 & $-$44 22 56.4 & 15\phd\phn\phn & 2 & 26 & [13] & 6.81 & [3] & 3.26 & [4] & 1.7\phn & [5] & 0.82 & [9] & 0.36 & [7] & $-$1.12 & 33 & m &  65 \\
MRC & B0008$-$421 & 00 10 52.55 & $-$41 53 10.3 & 2.69\rlap{:} & 3 & \nodata & \nodata & 6.61 & [3] & 6.52 & [1] & 4.62 & [14] & 2.49 & [14] & 1.32 & [14] & $-$0.52 & $\leq$10 & m & \nodata \\
 & & & & & & & & & & & & & & & & & & & & & \\
MRC & B0012$-$383 & 00 15 24.33 & $-$38 04 33.6 & 10.7\phn & 2 & 21 & [13] & 4.87 & [3] & 2.18 & [4] & 1.26 & [15] & 0.65 & [9] & 0.35 & [7] & $-$1.07 & 20 & m &  78 \\
MRC & B0013$-$634 & 00 16 02.59 & $-$63 10 06.6 & 15\phd\phn\phn & 2 & 27 & [2] & 6.90 & [3] & 3.07 & [4] & 1.9\phn & [11] & 0.82 & [6] & 0.45 & [7] & $-$1.13 & 31 & m &  76 \\
\enddata
\end{deluxetable}

\clearpage
\begin{deluxetable}{rrcll}
\tabletypesize{\footnotesize}
\tablecaption{Sample size, source density, and median values of various
quantities for sources in the MS4, SMS4, and 3CRR
samples.\label{tab6.medians}}
\tablehead{
\colhead{(1)} & \colhead{(2)} & \colhead{(3)} & \multicolumn{2}{c}{(4)} \\
\colhead{} & \colhead{} & \multicolumn{1}{c}{Source} &
\multicolumn{2}{c}{Median values} \\
\colhead{Sample} & \colhead{N} & \colhead{density} &
\colhead{$S_{178}$} & \colhead{$\alpha$} \\
\colhead{} & \colhead{} & \colhead{(sr$^{-1}$)} &
\colhead{(Jy)} & \colhead{}}
\startdata
 & & & & \\
MS4 & 228 & $94.0 \pm 6.2$ & \merr{12.1}{-0.7}{+0.6} &
  \merr{-0.86}{-0.02}{+0.03} \\
 & & & & \\
SMS4 & 137 & $56.5 \pm 4.8$ & \merr{15.7}{-0.7}{+0.5} &
  \merr{-0.91}{-0.02}{+0.04} \\
 & & & & \\
3CRR & 172 & $40.6 \pm 3.1$ & \merr{15.6}{-0.8}{+0.8} &
  \merr{-0.81}{-0.01}{+0.03} \\
 & & & & \\
\enddata
\end{deluxetable}

\end{document}